\documentclass[12pt, a4paper]{article}
\usepackage[dvipdfmx]{graphicx}
\usepackage{color}
\usepackage{amssymb}
\usepackage{amsmath}
\usepackage{hyperref}
\usepackage{authblk}
\usepackage[a4paper, total={16.5cm,22cm}]{geometry}
\usepackage{subcaption}
\usepackage[sort&compress,numbers, merge]{natbib}

\newcommand{\lrf}[2]{ \left(\frac{#1}{#2}\right) }
\newcommand{\lrfp}[3]{ \left(\frac{#1}{#2}\right)^{#3} }
\newcommand{\vev}[1]{\left\langle #1\right\rangle}
\newcommand\unit[1]{\,\mathrm{#1}}
\newcommand\eV{\unit{eV}}

\newcommand\GeV{\unit{GeV}}

\title{{\bfseries\Large
Leptogenesis in the minimal gauged U(1)$_{L_\mu-L_\tau}$ model\\
and the sign of the cosmological baryon asymmetry
}}
\author[1]{Kento Asai}
\author[1,2]{Koichi Hamaguchi}
\author[1]{Natsumi Nagata}
\author[1]{Shih-Yen Tseng}
\affil[1]{{\small \textit{Department of Physics, University of Tokyo, Tokyo 113--0033,Japan}}}
\affil[2]{{\small \textit{Kavli IPMU (WPI), UTIAS, The University of Tokyo, Kashiwa, Chiba 277--8583, Japan}}}
\date{}

\begin{document}
\begin{flushright}
 IPMU-20-0048 
\end{flushright}

{\let\newpage\relax\maketitle}

\begin{abstract}

The minimal gauged U(1)$_{L_\mu-L_\tau}$ model is a simple extension of the Standard Model and has a strong predictive power for the neutrino sector. In particular, the mass spectrum and couplings of heavy right-handed neutrinos are determined as functions of three neutrino Dirac Yukawa couplings, with which we can evaluate the baryon asymmetry of the Universe generated through their decay, \textit{i.e.}, leptogenesis. In this letter, we study leptogenesis in the minimal gauged U(1)$_{L_\mu-L_\tau}$ model. It turns out that the sign of the resultant baryon asymmetry for the case with the Dirac CP phase, $\delta$, larger than $\pi$ is predicted to be opposite to that for $\delta < \pi$. In addition, if lepton asymmetry is dominantly produced by the decay of the lightest right-handed neutrino, then the correct sign of baryon asymmetry is obtained for $\delta > \pi$, which is favored by the current neutrino-oscillation experiments, whilst the wrong sign is obtained for $\delta < \pi$. We further investigate a non-thermal leptogenesis scenario where the U(1)$_{L_\mu-L_\tau}$ breaking field plays the role of inflaton and decays into right-handed neutrinos, as a concrete example. It is found that this simple framework offers a successful inflation that is consistent with the CMB observation. We then show that the observed amount of baryon asymmetry can be reproduced in this scenario, with its sign predicted to be positive in most of the parameter space.

\end{abstract}

\thispagestyle{empty}
\clearpage

%%%%%%%%%%%%%%%%%%%%%%%%%
\section{Introduction}
%%%%%%%%%%%%%%%%%%%%%%%%%

The origin of the baryon asymmetry of the Universe is one of the fundamental puzzles in particle physics and cosmology. 
Leptogenesis~\cite{Fukugita:1986hr} provides a simple and elegant explanation to this puzzle. In this scenario, heavy right-handed neutrinos are introduced and play a double role; they explain the small neutrino masses via the seesaw mechanism~\cite{Minkowski:1977sc,Yanagida:1979as,GellMann:1980vs,Mohapatra:1979ia}, while their CP-violating decay in the early Universe leads to a lepton asymmetry, which is then partially converted to a baryon asymmetry through the Standard Model sphaleron processes~\cite{Kuzmin:1985mm}. The resultant amount of baryon asymmetry depends on the structure of the neutrino sector, and a specific model for the neutrino sector could give a distinct prediction for baryon asymmetry.

In this letter, we study leptogenesis in the minimal gauged U(1)$_{L_\mu-L_\tau}$ model~\cite{Foot:1990mn, He:1990pn, He:1991qd, Foot:1994vd}, where a new U(1) gauge symmetry, called the U(1)$_{L_\mu-L_\tau}$ gauge symmetry, is introduced to the Standard Model in addition to a scalar field that spontaneously breaks this gauge symmetry. Although this model is a simple extension of the Standard Model, it has rich phenomenological implications because of its strong predictive power for the neutrino sector~\cite{Branco:1988ex, Choubey:2004hn, Araki:2012ip, Heeck:2014sna, Crivellin:2015lwa, Plestid:2016esp, Asai:2017ryy, Asai:2018ocx, Asai:2019ciz}. In this model, the light neutrino mass matrix has the so-called two-zero minor structure~\cite{Araki:2012ip, Heeck:2014sna, Crivellin:2015lwa, Asai:2017ryy, Asai:2018ocx, Asai:2019ciz}, which allows us to determine the masses of light neutrinos, $m_i$ ($i=1,2,3$), as well as the three CP phases in the Pontecorvo-Maki-Nakagawa-Sakata (PMNS) mixing matrix~\cite{Pontecorvo:1967fh, Pontecorvo:1957cp, Pontecorvo:1957qd,Maki:1962mu}---the Dirac phase $\delta$ and the two Majorana phases $\alpha_{2,3}$---as functions of the neutrino mixing angles $\theta_{ij}$ and the squared mass differences $\Delta m_{ij}^2$~\cite{Asai:2017ryy, Asai:2018ocx, Asai:2019ciz}. Moreover, the mass spectrum and couplings of heavy right-handed neutrinos are also determined as functions of three neutrino Dirac Yukawa couplings.

Because of this restrictive structure of the neutrino sector and the small number of free parameters, we can thoroughly study leptogenesis in this model. This is the aim of the present work. We find that there is a close relationship between the Dirac CP phase and the sign of baryon asymmetry predicted in this model, as pointed out in Ref.~\cite{Asai:2017ryy}. In particular, the sign of baryon asymmetry obtained for $\delta > \pi $ is found to be opposite to that for $\delta < \pi $. As we will show, the Dirac CP phase favored by the current neutrino oscillation experiments, $\delta\sim 220^{\circ}$, naturally leads to the correct sign of baryon asymmetry in most cases.\footnote{The sign of the baryon asymmetry of the Universe in leptogenesis has been discussed in Refs.~\cite{Frampton:2002qc, Raidal:2002xf, Branco:2002xf, Pascoli:2003uh}.
}

To see this feature with a concrete example, we investigate a non-thermal leptogenesis scenario where the U(1)$_{L_\mu-L_\tau}$ breaking field is regarded as inflaton and it decays into right-handed neutrinos after inflation ends. These right-handed neutrinos are supposed to be out of thermal equilibrium, and their decay generates a lepton asymmetry non-thermally. It is found that this simple framework offers a successful inflation that is consistent with the CMB observation. We then show that the observed amount of baryon asymmetry can be reproduced and, in particular, its sign is predicted to be positive in a wide range of parameter space.

%%%%%%%%%%%%%%%%%%%%%%%%%%%%%%%%%%%%%%%%%%%%%%%%%%%%
\section{Minimal gauged U(1)$_{L_\mu-L_\tau}$ model}
\label{sec:model}
%%%%%%%%%%%%%%%%%%%%%%%%%%%%%%%%%%%%%%%%%%%%%%%%%%%%

Let us first briefly review the minimal gauged U(1)$_{L_\mu-L_\tau}$ model~\cite{Harigaya:2013twa, Kaneta:2016uyt, Asai:2017ryy, Asai:2018ocx, Nomura:2019uyz, Asai:2019ciz}. To successfully reproduce the neutrino mixing and be consistent with the constraints on the neutrino oscillation parameters, we introduce three heavy right-handed neutrinos and an SU(2)$_L$ singlet field to the Standard Model, denoted by $N_\alpha$ ($\alpha=e,\mu,\tau$) and $\sigma$, respectively. The U(1)$_{L_\mu-L_\tau}$ charges of the field content in this model are given by
\begin{align}
\begin{cases}
\mu_R, L_\mu, N_\mu: & +1,\\
\tau_R, L_\tau, N_\tau: & -1,\\
\sigma:& +1,\\
\mathrm{others}:&0,
\end{cases}
\end{align}
where $e_R,\mu_R,\tau_R$ are the right-handed charged leptons, and $L_\alpha$ ($\alpha = e, \mu, \tau$) are the left-handed lepton doublets. The scalar field $\sigma$ spontaneously breaks the U(1)$_{L_\mu-L_\tau}$ gauge symmetry when it develops a vacuum expectation value (VEV). With these fields, the most general renormalizable interaction terms in the neutrino sector are given by
\begin{align}
{\cal L}_N = 
&-\lambda_e N_e^c (L_e \cdot H)
-\lambda_\mu N_\mu^c (L_\mu \cdot H)
-\lambda_\tau N_\tau^c (L_\tau \cdot H) \nonumber \\
&-\frac{1}{2}M_{ee} N_e^c N_e^c 
- M_{\mu \tau} N_\mu^c N_\tau^c 
- \frac{1}{2} \sum_{\alpha,\beta = e,\mu} h_{\alpha \beta} \sigma N_{\alpha}^c N_{\beta}^c
- \frac{1}{2} \sum_{\alpha,\beta = e,\tau} h_{\alpha \beta} \sigma^{\ast} N_{\alpha}^c N_{\beta}^c +\text{h.c.} ~,
\label{eq:Lagrangian-1}
\end{align}
where the dots between $L$ and $H$ indicate the contraction of the SU(2)$_L$ indices, and $h_{\alpha\beta}$ ($\alpha,\beta = e,\mu,\tau$) is a symmetric matrix with $h_{e\mu} = h_{\mu e}$, $h_{e\tau} = h_{\tau e}$, and all other elements being zero.
After the Higgs field $H$ and the scalar field $\sigma$ acquire VEVs, $\vev{H}\simeq 174\GeV$ and $\vev{\sigma}$, respectively, these Lagrangian terms lead to a diagonal Dirac neutrino mass matrix and a Majorana mass matrix with a special structure of two zero components:
\begin{align}
 {\cal M}_D = 
\begin{pmatrix}
 \lambda_e & 0& 0\\
 0 & \lambda_\mu & 0 \\
 0 & 0 & \lambda_\tau 
\end{pmatrix}
 \vev{H}\,,
 \qquad
{\cal M}_R =
\begin{pmatrix}
 M_{ee} & h_{e\mu} \vev{\sigma} & h_{e\tau} \vev{\sigma} \\
h_{e\mu} \vev{\sigma} & 0 & M_{\mu\tau} \\
h_ {e\tau} \vev{\sigma} & M_{\mu\tau} & 0
\end{pmatrix}.
\label{eq:MDandMR}
\end{align}
We can take $\lambda_\alpha$ ($\alpha=e,\mu,\tau$) and $\vev{\sigma}$ to be real and positive via field redefinitions without loss of generality. 
The light neutrino mass matrix is then given by the seesaw formula~\cite{Minkowski:1977sc,Yanagida:1979as,GellMann:1980vs,Mohapatra:1979ia}, ${\cal M}_\nu = -{\cal M}_D {\cal M}_R^{-1} {\cal M}^{T}_D$, and since it is a complex symmetric matrix, it can be diagonalized with a unitary matrix as $U^T {\cal M}_{\nu} U =\text{diag}(m_1, m_2, m_3)$, where the unitary matrix $U$ is the PMNS mixing matrix and is parametrized as  
\begin{equation}
 U = 
\begin{pmatrix}
 c_{12} c_{13} & s_{12} c_{13} & s_{13} e^{-i\delta} \\
 -s_{12} c_{23} -c_{12} s_{23} s_{13} e^{i\delta}
& c_{12} c_{23} -s_{12} s_{23} s_{13} e^{i\delta}
& s_{23} c_{13}\\
s_{12} s_{23} -c_{12} c_{23} s_{13} e^{i\delta}
& -c_{12} s_{23} -s_{12} c_{23} s_{13} e^{i\delta}
& c_{23} c_{13}
\end{pmatrix}
\begin{pmatrix}
 1 & & \\
 & e^{i\frac{\alpha_{2}}{2}} & \\
 & & e^{i\frac{\alpha_{3}}{2}}
\end{pmatrix}
~,
\end{equation}
where $c_{ij} \equiv \cos \theta_{ij}$ and $s_{ij} \equiv \sin
\theta_{ij}$ with the mixing angles $\theta_{ij} \in [0, \pi/2]$, and the Dirac CP phase $\delta \in [0, 2\pi]$,
and the order $m_1<m_2$ is chosen without loss of generality. 
We follow the convention of the Particle Data Group~\cite{Tanabashi:2018oca},
where $0<\Delta m_{21}^2 \ll |\Delta m_{31}^2|$ with $\Delta m_{ij}^2 = m_i^2-m_j^2$. 

In the minimal gauged U(1)$_{L_\mu-L_\tau}$ model, because of the two-zero matrix structure in Eq.~\eqref{eq:MDandMR}, the light neutrino mass matrix is subject to the two-zero minor conditions~\cite{Lavoura:2004tu,Lashin:2007dm}: $[{\cal M}_\nu^{-1}]_{\mu\mu} = [{\cal M}_\nu^{-1}]_{\tau\tau} = 0$.\footnote{This two-zero minor structure of ${\cal M}_\nu$ is stable against the renormalization group effect, as shown in Ref.~\cite{Asai:2017ryy}.} These two complex equations impose four constraints on the parameters. As a result, among the nine degrees of freedom in the light neutrino sector, the lightest neutrino mass $m_{0}$ and the three CP phases $\delta$, $\alpha_2$ and $\alpha_3$ are uniquely determined as functions of the other five parameters, $\theta_{ij}$ and $\Delta m_{ij}^2$, which are observables in neutrino oscillation experiments~\cite{Asai:2017ryy}:
\begin{align}
m_{0} = m_{0}(\theta_{ij}, \Delta m_{ij}^2)~,\;
\delta = \delta(\theta_{ij}, \Delta m_{ij}^2)~,\;
\alpha_{2, 3} = \alpha_{2, 3}(\theta_{ij}, \Delta m_{ij}^2)~.
\label{eq:m0delalp}
\end{align}
As shown in Ref.~\cite{Asai:2017ryy}, the case for inverted hierarchy does not work, and hence we concentrate on the case of the normal hierarchy in the following analysis.

%%%
\begin{figure}[t]
\centering
\subcaptionbox{\label{fig:delta} Dirac CP phase}{
\includegraphics[width=0.45\columnwidth]{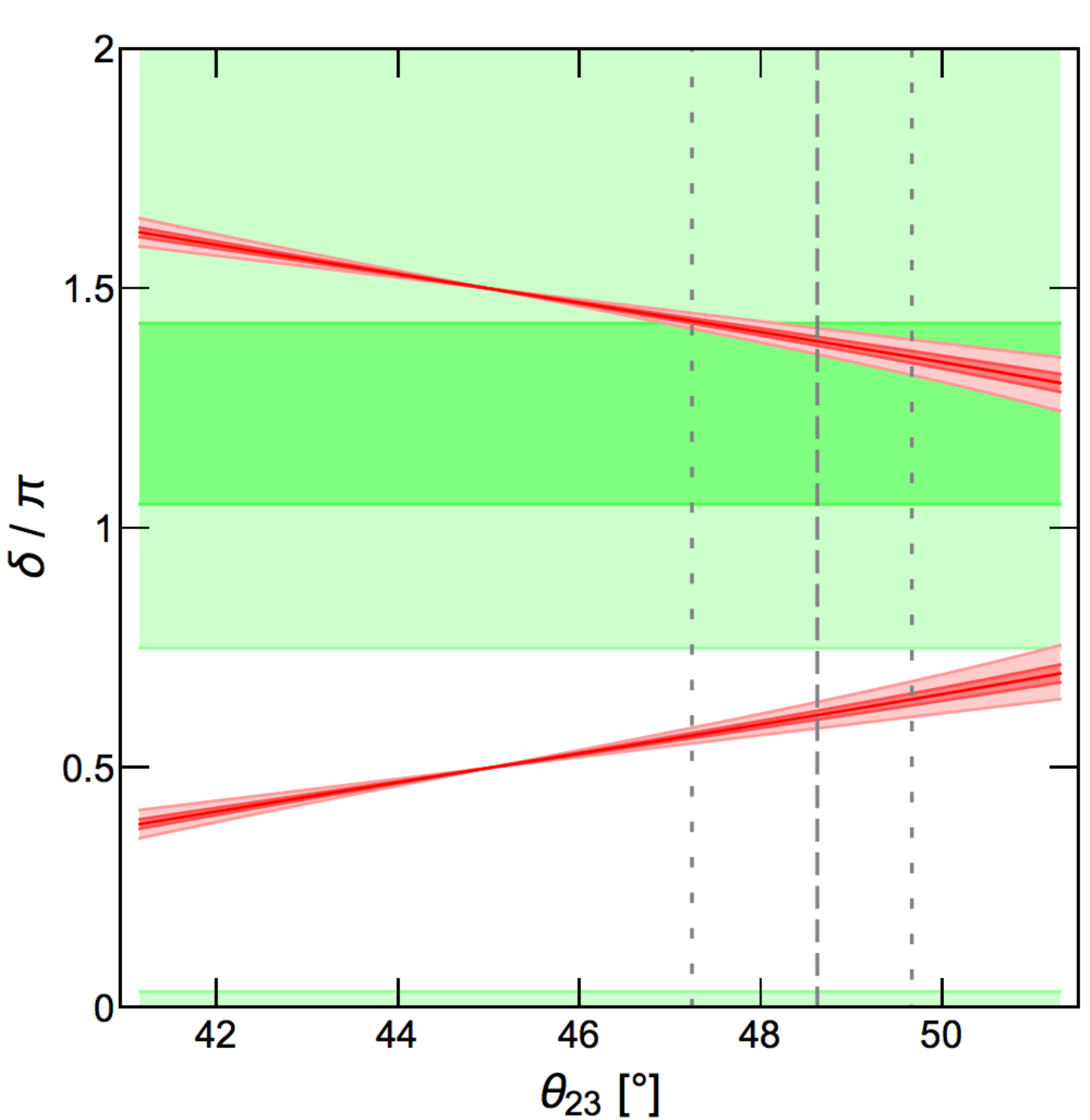}}
\hspace{5mm}
\subcaptionbox{\label{fig:sum} Sum of the neutrino masses}{
\includegraphics[width=0.45\columnwidth]{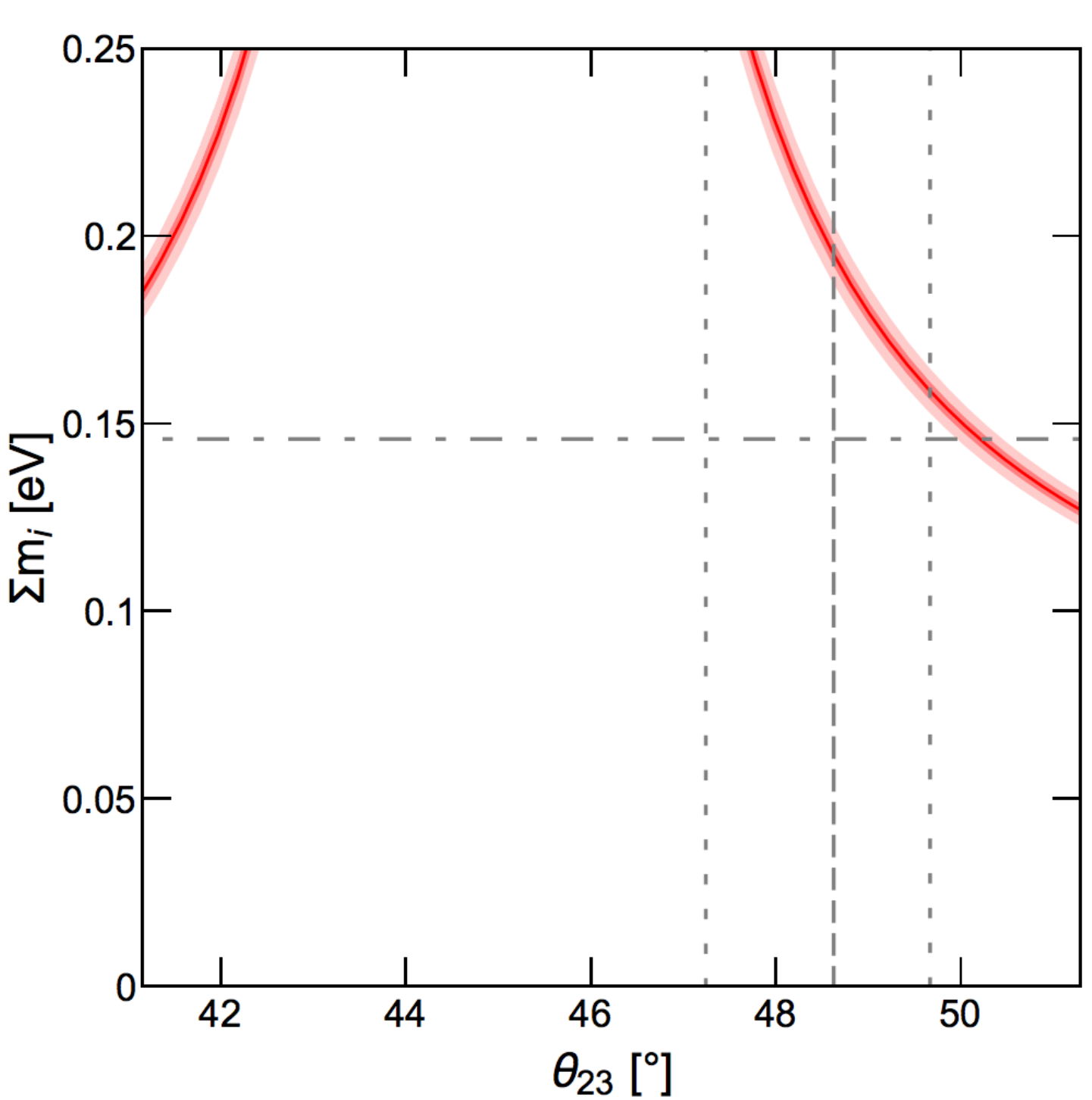}}
\caption{The Dirac CP phase $\delta$ and the sum of the neutrino masses $\sum_i m_i$ as a function of $\theta_{23}$. $\theta_{23}$ is varied in the $3\sigma$ range, and the $1\sigma$ range is in between the vertical dotted lines, with the vertical dashed line showing the central value. The dark (light) red bands show the uncertainty coming from the $1\sigma$ ($3\sigma$) errors in the parameters $\theta_{12}$, $\theta_{13}$, $\Delta m_{21}^2$, and $\Delta m_{31}^2$. The dark (light) horizontal green band in Fig.~(a) is the $1\sigma$ ($3\sigma$) favored region of $\delta$. The horizontal dot-dashed line in Fig.~(b) is the present limit given by Ref.~\cite{RoyChoudhury:2019hls}: $\sum_i m_i < 0.146\eV$ (normal ordering, 95\% C.L.).}
\label{fig:delta_and_sum}
\end{figure}
%%%

In Fig.~\ref{fig:delta_and_sum}, we show the Dirac CP phase $\delta$ and the sum of the light neutrino masses $\sum_i m_i$ as a function of $\theta_{23}$.
Fig.~\ref{fig:delta} shows the Dirac CP phase, where the dark (light) red bands show the uncertainty coming from the other parameters, $1\sigma$ ($3\sigma$) errors in the parameters $\theta_{12}$, $\theta_{13}$, $\Delta m_{21}^2$, and $\Delta m_{31}^2$, adopted from the NuFIT 4.1 global analysis of neutrino oscillation measurements~\cite{Esteban:2018azc, NuFIT4_1}. Note that there are two solutions for Dirac CP phase, $\delta$ and $2\pi -\delta$. Currently, $\delta>\pi$ is favored experimentally; in particular, the latest result given by the T2K Collaboration sets the 3$\sigma$ confidence interval for $\delta$ as [$0.915\pi$--$1.99\pi$] for the normal-ordering case~\cite{Abe:2019vii}. The prediction of the sum of the light neutrino masses is shown in Fig.~\ref{fig:sum}, together with the cosmological constraint for the normal ordering case, $\sum_i m_i < 0.146\eV$ (95\% C.L.)~\cite{RoyChoudhury:2019hls}.\footnote{The analysis in Ref.~\cite{RoyChoudhury:2019hls} takes account of the neutrino mass-squared splittings, which results in a more conservative bound than that set by the Planck experiment, 
$\sum_i m_i < 0.12\eV~(95\%~\text{C.L.})$~\cite{Aghanim:2018eyx}, obtained on the assumption of degenerate neutrino masses. See also Ref.~\cite{Ivanov:2019hqk}, where a similar conclusion is drawn with a more relaxed bound: $\sum_i m_i < 0.16\eV~(95\%~\text{C.L.})$. } As can be seen in this figure, the model is driven into a corner~\cite{Asai:2018ocx}, but the region around $\theta_{23}\sim 50^\circ$ is still marginally viable.

In the numerical analysis performed in the subsequent sections, as a benchmark point, we adopt the following values of the neutrino oscillation parameters for the normal ordering case: 
\begin{align}
&\sin^2 \theta_{12} = 0.310, \quad \sin^2\theta_{23} = 0.604, \quad \sin^2 \theta_{13}=0.0224,
\nonumber \\
&\Delta m_{21}^2 = 7.39\times 10^{-5}\eV^2, \quad \Delta m_{31}^2 = 2.53\times 10^{-3}\eV^2,
\label{eq:oscillation_parameters}
\end{align}
where we use the central values of $\theta_{12}$, $\theta_{13}$, and $\Delta m_{ij}^2$  obtained in the NuFIT 4.1 global analysis of neutrino oscillation measurements~\cite{Esteban:2018azc, NuFIT4_1}, while for $\theta_{23}$ we set $\theta_{23}=51^\circ$ in order to evade the cosmological constraint on the sum of light neutrino masses. At this benchmark point, the other neutrino parameters are predicted to be $m_1=m_0\simeq 3.45\times 10^{-2}$~eV, $\delta\simeq 236^\circ$, and $(\alpha_1,\alpha_2)\simeq (-123^\circ,84.3^\circ)$, where between the two-fold degenerate solutions of $\delta$ we have chosen the one with $\delta>\pi$. Another interesting observable is the effective Majorana mass $\langle m_{\beta\beta} \rangle$, which is defined by
\begin{align}
\langle m_{\beta\beta} \rangle \equiv \biggl|
\sum_{i} U_{ei}^2 m_i
\biggr|
=\left|
c_{12}^2 c_{13}^2 m_1 +
s_{12}^2 c_{13}^2 e^{i \alpha_2} m_2 +
s_{13}^2 e^{i(\alpha_3 -2\delta)} m_3 
\right| ~.
\label{eq:double_beta}
\end{align}
The neutrinoless double-beta decay rate is proportional to $\vev{m_{\beta\beta}}^{2}$. For the above benchmark point, we obtain $\langle m_{\beta\beta} \rangle = 0.021$~eV. This is well below the current constraint given by the KamLAND-Zen experiment, $\langle m_{\beta\beta} \rangle < 0.061$--0.165 eV~\cite{KamLAND-Zen:2016pfg}, where the uncertainty of this upper bound is due to the error in the nuclear matrix element of $^{136}$Xe. Future experiments are expected to be sensitive to $\langle m_{\beta\beta} \rangle = \mathcal{O}(0.01)$~eV \cite{Agostini:2017jim,Agostini:2020adk} and thus potentially able to test this prediction.

%%%%%%%%%%%%%%%%%%%%%%%%%%%%%%%%%%%%%%%%%%%%%%%%%%%%%%%%%%%%%%%%%%
\section{Asymmetry parameter of the right-handed neutrino decay}
\label{sec:asympar}
%%%%%%%%%%%%%%%%%%%%%%%%%%%%%%%%%%%%%%%%%%%%%%%%%%%%%%%%%%%%%%%%%%

Next, we discuss the CP-violating decay of heavy right-handed neutrinos in our model. As seen above, the effective light neutrino mass matrix is subject to the two-zero minor conditions, \textit{i.e.}, the inverse of the matrix, ${\cal M}^{-1}_\nu$, contains two zeros among its nine components. These two conditional equations force four free parameters to be dependent on the others, whose values are fixed in the following analysis as in Eq.~\eqref{eq:oscillation_parameters}. Still, there are several coupling constants undetermined in the Lagrangian in Eq.~\eqref{eq:Lagrangian-1}. It is then convenient to take $\lambda_\alpha$ as input parameters; with this choice, all the entries in ${\cal M}_D$ and ${\cal M}_R$ are uniquely determined in terms of $\lambda_\alpha$ and the neutrino oscillation parameters. By diagonalizing the mass matrix of the right-handed neutrinos, the Lagrangian \eqref{eq:Lagrangian-1} can be rewritten as
\begin{align}
{\Delta \cal L} = 
&-\hat{\lambda}_{i\alpha} \hat{N}_{i}^{c} (L_{\alpha} \cdot H) -\frac{1}{2} M_{i} \hat{N}_{i}^{c} \hat{N}_{i}^{c} + \text{h.c.}~,
\label{eq:Lagrangian_MR_diag}
\end{align}
where
\begin{align}
{\cal M}_R &= \Omega^{\ast} \text{diag}(M_1,M_2,M_3) \Omega^{\dagger}~, \\
\hat{N}_{i}^{c} &= \sum_{\alpha} \Omega_{\alpha i}^{\ast} N_{\alpha}^{c}~, \\
\hat{\lambda}_{i\alpha} &= \Omega_{\alpha i} \lambda_{\alpha}~(\text{not~summed})~,
\end{align}
where $\Omega $ is a unitary matrix and $M_i$ ($i=1,2,3$) are the mass eigenvalues of ${\cal M}_R$. These quantities are, again, uniquely determined in terms of $\lambda_\alpha$, for a given set of the neutrino oscillation parameters. 

In leptogenesis, the final baryon asymmetry depends on the asymmetry parameters of the decay of right-handed neutrinos:\footnote{
In the following analysis, we assume that the mass of the U(1)$_{L_\mu-L_\tau}$ gauge boson, $Z^\prime$, is sufficiently large so that the decay modes that contain $Z^\prime$ in the final state, such as $\hat{N}_i \to \hat{N}_j Z^\prime$ and $\hat{N}_i \to \nu_j Z^\prime$, are kinematically forbidden. }
\begin{align}
\epsilon_i = \frac{\Gamma(\hat{N}_i\to L H)-\Gamma(\hat{N}_i\to \bar{L} H^*)}{\Gamma(\hat{N}_i\to L H)+\Gamma(\hat{N}_i\to \bar{L} H^*)}~.
\end{align}
At the leading order, it is computed as~\cite{Flanz:1994yx,Covi:1996wh,Buchmuller:1997yu}
\begin{align}
\epsilon_{i} &= \frac{1}{8\pi} \frac{1}{ \big( \hat{\lambda} \hat{\lambda}^{\dagger} \big)_{ii}} \sum_{j\neq i} \mathrm{Im} \big\{ \big( \hat{\lambda} \hat{\lambda}^{\dagger} \big)^{2}_{ij}  \big\} f\bigg(\frac{M_{j}^{2}}{M_{i}^{2}}\bigg)~, \label{eq:epsilon_1} \\
f(x) &= \sqrt{x} \bigg[ 1 - (1+x)\mathrm{ln}\bigg(\frac{1+x}{x}\bigg) + \frac{1}{1-x} \bigg]~.
\label{eq:epsilononeloop}
\end{align}
The significance of the effect of each asymmetry parameter on the resultant lepton asymmetry highly depends on the leptogenesis scenarios.  
In the thermal leptogenesis, for instance, the decay of the lightest right-handed neutrino tends to give the dominant contribution to the lepton asymmetry, since the asymmetry generated by the heavier right-handed neutrinos are washed out---in this case, the final baryon asymmetry, $n_B$, is essentially proportional to $\epsilon_1$. In the case discussed in the next section, we will consider the decay of all three right-handed neutrinos.

In the present scenario, the sign of $\epsilon_i$, and thus that of the resultant baryon asymmetry as well, for the case of $\delta > \pi$ turns out to be opposite to that for $\delta < \pi$~\cite{Asai:2017ryy}. As can easily be seen from the analytical expressions for the Majorana CP phases $\alpha_{2,3}$ given in Ref.~\cite{Asai:2017ryy}, the transformation $\delta \to 2\pi - \delta$ leads to $U \to U^*$, which then results in ${\cal M}_\nu \to {\cal M}_\nu^*$, ${\cal M}_R \to {\cal M}_R^*$, $\Omega \to \Omega^*$, $\hat{\lambda} \to \hat{\lambda}^*$, and thus $\epsilon_i \to - \epsilon_i$. In order to obtain the correct sign (positive) for baryon asymmetry in leptogenesis, the generated lepton asymmetry $n_L$ must be negative, because the sphaleron processes predict $n_{B}/n_{L} < 0$~\cite{Harvey:1990qw}. This, in particular, indicates that $\epsilon_1$ should be negative when the decay of the lightest right-handed neutrino predominantly generates lepton asymmetry.

%%%%%%%%%%%%%%%%%%%%%%%%%%%%%%%%%%%%%%%%%%%%%%%%%%%%%%%%%%%%%%
\begin{figure}[t]
\centering
\includegraphics[width=0.8\columnwidth]{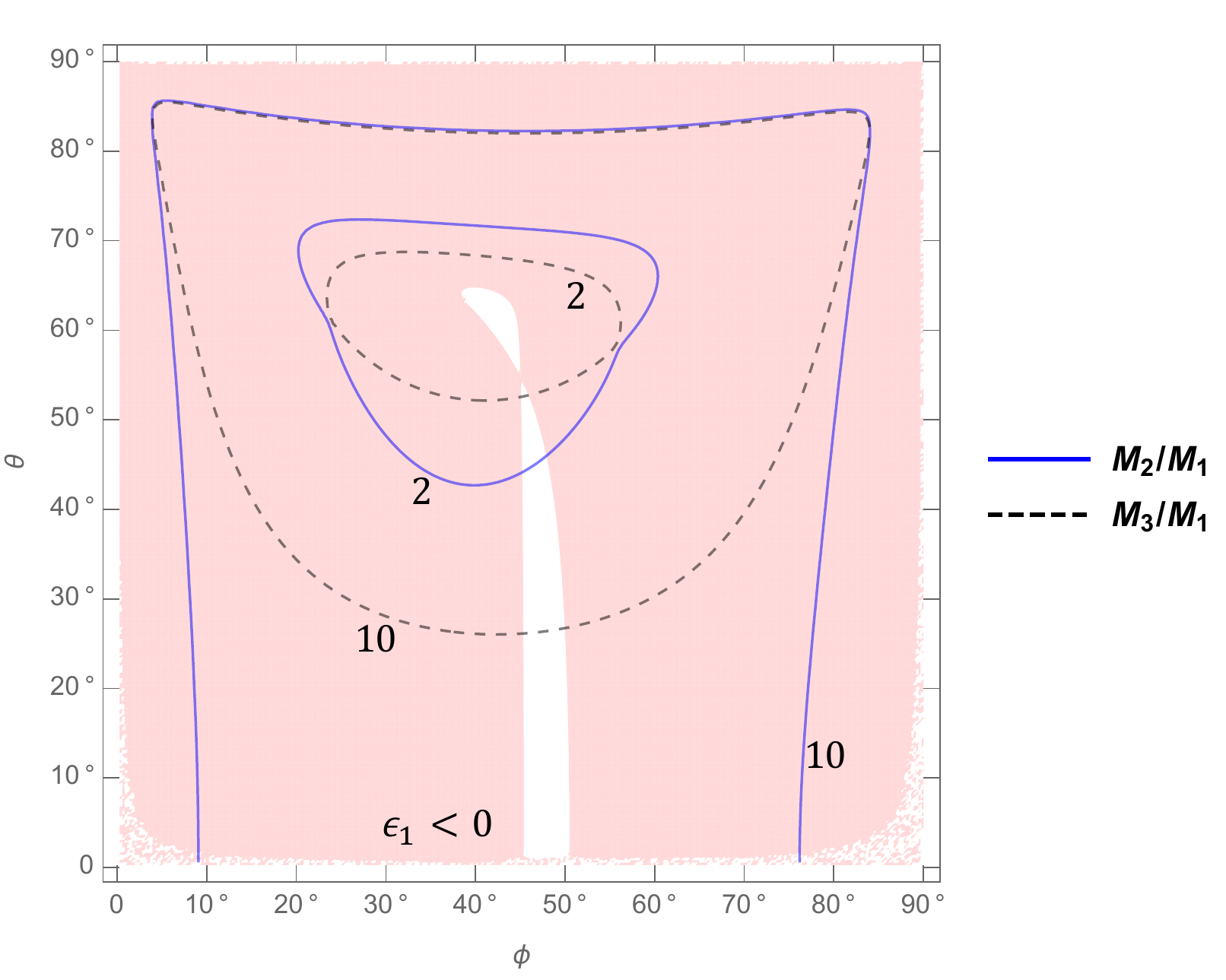}
\caption{The sign of the asymmetry parameter for the lightest right-handed neutrino, $\epsilon_1$, in the $\phi$-$\theta$ plane for $\delta>\pi$. The shaded region corresponds to a negative value of $\epsilon_1$. For $\delta<\pi$, the sign of $\epsilon_1$ is flipped. The blue solid and gray dashed contour lines show the ratios of the right-handed neutrino masses, $M_2/M_1$ and $M_3/M_1$, respectively. }
\label{fig:epsilon_1}
\end{figure}
%%%%%%%%%%%%%%%%%%%%%%%%%%%%%%%%%%%%%%%%%%%%%%%%%%%%%%%

To see the predicted sign of $\epsilon_1$ in our scenario, in Fig.~\ref{fig:epsilon_1}, we show the sign of the asymmetry parameter for the lightest right-handed neutrino, $\epsilon_1$. In visualizing this, we parametrize the coupling constants $\lambda_\alpha$ as
\begin{equation}
    (\lambda_e,\lambda_\mu,\lambda_\tau)=\lambda\,(\cos\theta, \sin\theta\cos\phi,\sin\theta\sin\phi) ~,
    \label{eq:lamthphi}
\end{equation}
with $0 \leq \theta,\phi \leq \pi/2$, and show $\text{sgn}(\epsilon_1)$ in the $\phi$-$\theta$ plane. Here, we take $\delta>\pi$, as favored by the neutrino oscillation data~\cite{Esteban:2018azc, NuFIT4_1}. We see that the desirable sign, $\epsilon_{1}<0$, is obtained in almost all the parameter region, except in the the small crack around $\phi=45^{\circ}$. Notice that, as discussed above, the sign of $\epsilon_1$ is flipped under the transformation $\delta \to -\delta + 2\pi$. This has interesting implications for our model; the experimentally favored Dirac CP phase, $\delta>\pi$, generically leads to the correct sign of baryon asymmetry, whilst the experimentally disfavored one, $\delta < \pi$, yields the wrong sign.\footnote{Note that the sign of $\epsilon_1$ obtained here is opposite to that found in Ref.~\cite{Asai:2017ryy}. This difference is attributed to the different choices of the input parameters, especially $\theta_{23}$; in Ref.~\cite{Asai:2017ryy}, the value of $\theta_{23}$ was taken from the global fit performed in Ref.~\cite{Capozzi:2017ipn}, which was in the first octant---the preferred value of $\theta_{23}$ has moved to the second octant since then, as found in Refs.~\cite{Esteban:2018azc, NuFIT4_1}, and this change results in a sign flip in $\epsilon_1$. }

We also show in Fig.~\ref{fig:epsilon_1} the ratios of the right-handed neutrino masses, $M_2/M_1$ and $M_3/M_1$, in the blue solid and gray dashed contour lines, respectively. It is found that this model predicts a moderately degenerate mass spectrum for right-handed neutrinos, except near the edge of the plane.

As can be seen from Eq.~\eqref{eq:epsilononeloop}, the asymmetry parameters $\epsilon_i$ are proportional to $\lambda^2$. We have also checked that the magnitude of the asymmetry parameter $\epsilon_1$ is predicted to be $|\epsilon_1|/\lambda^2 \lesssim {\cal O}(10^{-4})$ in the typical parameter region in Fig.~\ref{fig:epsilon_1}, and thus the observed baryon asymmetry can be reproduced for a sufficiently large $\lambda$. A more detailed analysis for the thermal leptogenesis in the present scenario is beyond the scope of this letter and will be given elsewhere. In the next section, instead, we study the non-thermal leptogenesis for a minimal inflation scenario in our model.

%%%%%%%%%%%%%%%%%%%%%%%%%%%%%%%%%%%%%%%%%%%%%%%%%%%%%%%%%%%%%%%%%%%%%
\section{Inflation and non-thermal leptogenesis in the minimal gauged U(1)$_{L_\mu-L_\tau}$ model}
%%%%%%%%%%%%%%%%%%%%%%%%%%%%%%%%%%%%%%%%%%%%%%%%%%%%%%%%%%%%%%%%%%%%%

%%%%%%%%%%%%%%%%%%%%%%%%%%%%%%%%%%%
\subsection{Inflation model}
\label{sec:inflation}
%%%%%%%%%%%%%%%%%%%%%%%%%%%%%%%%%%%%

Now we investigate the non-thermal leptogenesis that proceeds through the inflaton decay into right-handed neutrinos \cite{Lazarides:1991wu, Kumekawa:1994gx, Lazarides:1999dm, Giudice:1999fb, Asaka:1999yd, Asaka:1999jb, Kawasaki:2000ws, Hamaguchi:2002vc} in the minimal gauged U(1)$_{L_\mu-L_\tau}$ model. We identify the U(1)$_{L_\mu-L_\tau}$ breaking field $\sigma$ as the inflaton field and assume the following form of the Lagrangian terms for this field~\cite{Ema:2016ops} (see also Refs.~\cite{Kallosh:2013hoa, Ferrara:2013rsa, Kallosh:2013yoa, Galante:2014ifa}):
\begin{align}
{\cal L}_\sigma &=
\frac{|D_\mu\sigma|^2}{(1-|\sigma|^2/\Lambda^2)^2} - \kappa (|\sigma|^2-\vev{\sigma}^2)^2~,
\label{eq:lagsigma}
\end{align}
where $\Lambda$ is a parameter with mass dimension one, taken such that $\Lambda > \vev{\sigma}$. By using a U(1)$_{L_\mu-L_\tau}$ gauge transformation, we can always take the direction of the field excursion to be real; in this basis, $\varphi \equiv \sqrt{2} \text{Re} (\sigma)$ plays the role of the inflaton. 
The pole of the kinetic term guarantees that the effective potential becomes very flat at large field values~\cite{Takahashi:2010ky, Nakayama:2010kt}. The inflaton field is canonically normalized with a change of variable,
\begin{align}
\frac{\varphi}{\sqrt{2}\Lambda} &\equiv \tanh \lrf{\widetilde{\varphi}}{\sqrt{2}\Lambda}~,
\end{align}
which leads to 
\begin{equation}
    {\cal L}_\sigma 
= \frac{1}{2}\left(\partial_\mu \widetilde{\varphi}\right)^2-V(\widetilde{\varphi}) ~,
\end{equation}
with
\begin{align}
V(\widetilde{\varphi}) &= \kappa \Lambda^4 \left[\tanh^2\lrf{\widetilde{\varphi}}{\sqrt{2}\Lambda}-\lrfp{\vev{\sigma}}{\Lambda}{2}
\right]^2 ~.
\label{eq:inflatonpot}
\end{align}
This potential becomes flat for a large field value of $\widetilde{\varphi}$, allowing $\widetilde{\varphi}$ to behave as an inflaton field. As we see below, in the parameter region of interest, $\langle \sigma \rangle \ll \Lambda$; in this case, the VEV of the canonically-normalized field $\widetilde{\varphi}$ is simply given by $\langle \widetilde{\varphi} \rangle \simeq \sqrt{2} \langle \sigma \rangle$. Near this minimum, $\varphi$ differs from $\widetilde{\varphi}$ by a factor of $1-|\langle \sigma \rangle|^2/\Lambda^2$, which is very close to unity when $\langle \sigma \rangle \ll \Lambda$---we, thus, ignore this factor in the following expressions.

The scalar potential receives quantum corrections via the couplings of the $\sigma$ field with the right-handed neutrinos and the U(1)$_{L_\mu-L_\tau}$ gauge field.\footnote{The radiative corrections by the self coupling $\kappa$ is insignificant as long as $\kappa$ is perturbative.} These corrections turn out to be negligible if 
\begin{align}
    |h_{e\mu}|^2 + |h_{e\tau}|^2 & \ll 4\pi \sqrt{\kappa} ~, \label{eq:hllkap} \\
    g^2_{Z^\prime} &\ll 4\pi \sqrt{\kappa} ~, \label{eq:gzllkap}
\end{align}
where $g_{Z^\prime}$ is the U(1)$_{L_\mu-L_\tau}$ gauge coupling constant. We assume these conditions to be satisfied in the following analysis. 

The number of $e$-folds after the CMB modes left the horizon is defined by
\begin{equation}
    N_e \equiv \ln \biggl(\frac{a_f}{a_k}\biggr)~,
\end{equation}
where $a_f$ is the scale factor at the end of inflation, $a_k \equiv k/H_{\rm inf}$ with $H_{\rm inf}$ the Hubble parameter during inflation, and $k$ is a wave-number which corresponds to the CMB scale. We set $k$ equal to the default pivot scale adopted by the Planck collaboration~\cite{Aghanim:2018eyx}, $k = 0.05~\text{Mpc}^{-1}$, and evaluate $N_e$ as~\cite{Liddle:2003as}
\begin{align} \label{eq:efolding}
N_e &\simeq 62 + \frac{1}{3}\ln\lrf{H_{\rm{inf}}T_{R}}{M^{2}_{P}} \nonumber \\[2pt]
&\simeq 49 +\frac{1}{3}\ln \biggl(
\frac{H_{\rm inf}}{10^{11}~{\rm GeV}}
\biggr)+\frac{1}{3}\ln \biggl( 
\frac{T_R}{10^{9}~{\rm GeV}}
\biggr)
~,
\end{align}
where $T_R$ is the reheating temperature and $M_P=(8\pi G)^{-1/2} \simeq 2.4 \times 10^{18}$ GeV is the reduced Planck scale with $G$ being the gravitational constant. On the other hand, given the inflaton potential \eqref{eq:inflatonpot}, we can express $N_e$ in terms of the inflaton field $\widetilde{\varphi}$ as
\begin{align}
N_{e} &\simeq \int^{\widetilde{\varphi}_{N}}_{\widetilde{\varphi}_{f}} \lrf{V}{M^{2}_{P} V'} d\widetilde{\varphi} \nonumber \\
          &=\frac{1}{8M^{2}_{P}} \Bigg\{ \big(\Lambda^{2} - \vev{\sigma}^{2}\big)~\text{cosh}\lrf{2\widetilde{\varphi}}{\sqrt{2}\Lambda} - 4\vev{\sigma}^{2}~\text{ln} \left[\text{sinh}\lrf{\widetilde{\varphi}}{\sqrt{2}\Lambda} \right] \Bigg\} \bigg\rvert^{\widetilde{\varphi}_{N}}_{\widetilde{\varphi}_{f}}~,
          \label{eq:ne}
\end{align}
where $\widetilde{\varphi}_{N}$ and $\widetilde{\varphi}_{f}$ are the field values when the fluctuations observed in the CMB are created and inflation ends, respectively.  In the present model, inflation ends when $|V^{\prime \prime} M_P^2/V| \sim 1$
and it turns out that the corresponding field value $\widetilde{\varphi}_f$
is in general much smaller than $\widetilde{\varphi}_{N}$. 
In this case, we can obtain an approximate solution of Eq.~\eqref{eq:ne} with respect to $\widetilde{\varphi}_N$, by noting that $\langle \sigma \rangle \ll \Lambda$ and that the first term dominates the second term in Eq.~\eqref{eq:ne} for $\widetilde{\varphi}_N > \Lambda$:
\begin{align}
\widetilde{\varphi}_{N} \simeq \frac{\Lambda}{\sqrt{2}}~ \ln{\lrf{16N_{e}M^{2}_{P}}{\Lambda^{2}}}~.
\end{align}
We then evaluate the slow-roll parameters as
\begin{align}
    \epsilon &\equiv \frac{M_P}{2} \biggl(\frac{V^\prime}{V}\biggr)^2 
    \simeq \biggl(\frac{\Lambda}{2N_e M_P}\biggr)^2 ~,  \\
    \eta &\equiv \frac{V^{\prime \prime}}{V}M^{2}_{P}
     \simeq - \frac{1}{N_e} ~,
\end{align}
as well as the scalar spectral index $n_{s}$ and the tensor-to-scalar ratio $r$ as
\begin{align}
n_{s} &= 1 - 6\epsilon + 2\eta  \simeq 1-\frac{2}{N_{e}}~,\label{eq:ns} \\
r &= 16\epsilon \simeq \lrf{2\Lambda}{M_{P} N_{e}}^{2}
\simeq 3 \times 10^{-8} \times \biggl(\frac{\Lambda}{10^{16}~{\rm GeV}}\biggr)^2\biggl(\frac{N_e}{50}\biggr)^{-2}
~.
\end{align}
From Eq.~\eqref{eq:ns}, we see that $n_s \simeq 0.96$ for $N_e \simeq 50$; this is compatible with the Planck best-fit value $n_{s} = 0.9649 \pm 0.0042$~\cite{Aghanim:2018eyx}. On the other hand, the predicted value of the tensor-to-scalar ratio is much smaller than the Planck limit~\cite{Aghanim:2018eyx} and unable to be probed in the next-generation CMB experiments.

The power spectrum of the curvature perturbation $P_{\zeta}$ is 
\begin{align}
P_{\zeta} = \frac{V^{3}}{12\pi^{2}M^{6}_{P}V'^{2}} \nonumber \simeq \frac{\kappa N_e^2 \Lambda^2}{6\pi^{2} M_P^2}  ~.
\end{align}
With the measured value of the power spectrum, $P_{\zeta} \simeq (2.10\pm 0.03)\times 10^{-9}$~\cite{Aghanim:2018eyx}, we determine the coupling $\kappa$:
\begin{align}\label{eq:kappa}
\kappa \simeq 3\times 10^{-6} \times \lrf{N_e}{50}^{-2}\lrf{\Lambda}{10^{16}~\GeV}^{-2} ~.
\end{align}
We then obtain the Hubble parameter during inflation and the inflaton mass as 
\begin{align}\label{eq:Hinf}
H_{\text{inf}} &\simeq  \frac{\Lambda^2}{M_P}\sqrt{\frac{\kappa}{3}} \simeq 4 \times 10^{10}~\GeV \times \lrf{\Lambda}{10^{16} ~\GeV}
\biggl(\frac{N_e}{50}\biggr)^{-1}~, \\[2pt]
m_{\varphi} &\simeq  2 \sqrt{\kappa} \vev{\sigma} 
\simeq 3  \times 10^{10}~\mathrm{GeV} \times \lrf{\vev{\sigma}}{10^{13}~{\rm GeV}} \lrf{\Lambda}{10^{16} ~\GeV}^{-1} \biggl(\frac{N_e}{50}\biggr)^{-1} ~.
\label{eq:mphi}
\end{align}
In addition, the mass of the U(1)$_{L_\mu- L_\tau}$ gauge boson is given by 
\begin{equation}
    m_{Z^\prime} \simeq \sqrt{2} g_{Z^\prime} \langle \sigma \rangle ~.
\end{equation}

For $\Lambda \gg \langle \sigma \rangle$, the potential height at the origin is much lower than the potential energy during inflation. In this case, the U(1)$_{L_\mu- L_\tau}$ gauge symmetry would be restored during the (p)reheating process~\cite{Kofman:1995fi}. The subsequent symmetry breaking then leads to the formation of a cosmic-string network. Throughout cosmic history, oscillating string loops in the network emit gravitational waves, yielding a stochastic background of gravitational waves. The most stringent limits on this signature are imposed by pulsar timing arrays (PTAs), such as the Parkes PTA~\cite{Shannon:2015ect, Lasky:2015lej}, the European PTA~\cite{Lentati:2015qwp}, and the North American Nanohertz Observatory for Gravitational Waves~\cite{Arzoumanian:2015liz, Arzoumanian:2018saf}. With these data, as well as the theoretical predictions given in Refs.~\cite{Blanco-Pillado:2017rnf, Ringeval:2017eww}, one obtains $G\mu \lesssim \mathcal{O}(10^{-11})$, 
where $\mu$ is the mass per unit length of the cosmic string. For the Bogomol'nyi-Prasad-Sommerfield strings, which correspond to the case with $m_{\varphi} = m_{Z^\prime}$, we have $\mu = 2 \pi \vev{\sigma}^{2}$, for which the above limit leads to $\langle \sigma \rangle \lesssim 2 \times 10^{13}$~GeV. 
This bound slightly depends on the choice of parameters, $\kappa$ and $g_{Z^\prime}$, through the change in $\mu$. For example, in the limit $m_{Z^\prime} \gg m_{\varphi}$, we have $\mu \to 2 \pi \vev{\sigma}^{2}/ \ln(m_{Z^\prime}/m_{\varphi})$ \cite{Yung:1999du}, with which we obtain a weaker bound on $\langle \sigma \rangle$ than the aforementioned one. Future interferometric gravitational-wave detectors are expected to be sensitive to a much smaller value of $G\mu$; for example, the Laser Interferometer Space Antenna (LISA) can probe the gravitational waves emitted by cosmic strings with $G\mu \gtrsim 10^{-17}$~\cite{Auclair:2019wcv}, which corresponds to $\langle \sigma \rangle \gtrsim 2 \times 10^{10}$~GeV.

%%%%%%%%%%%%%%%%%%%%%%%%%%%%%%%%%%%%%%%%%%%%%%%%%%%%
\subsection{Reheating and non-thermal leptogenesis}
%%%%%%%%%%%%%%%%%%%%%%%%%%%%%%%%%%%%%%%%%%%%%%%%%%%%

After inflation ends, the Universe is reheated through the inflaton decay. In the following discussions, we consider the case where the inflaton decays dominantly into right-handed neutrinos; more specifically, we assume that the quartic coupling $\lambda_{H\sigma} |H|^2|\sigma|^2$ is negligibly small and that $m_{\varphi} < 2 m_{Z^\prime}$.\footnote{We can instead assume that $g_{Z}^\prime$ is negligibly small. In either case, we can always find a value of $g_{Z}^\prime$ that satisfies the condition \eqref{eq:gzllkap}.}  The total decay rate of inflaton in this case is given by
\begin{align}
 \Gamma_{\varphi} =\sum_{i, j} \frac{m_{\varphi}}{32 \pi}
&
\biggl[
1 - \frac{2(M_i^2+M_j^2)}{m_{\varphi}^2}
+ \frac{(M_i^2 -M_j^2)^2}{m_{\varphi}^4}
\biggr]^{\frac{1}{2}}
 \nonumber \\
&\times 
\biggl[
{\rm Re}\bigl(\hat{h}_{ij}\bigr)^2 
\biggl\{
1- \frac{(M_i+M_j)^2}{m_{\varphi}^2}
\biggr\}
+{\rm Im}\bigl(\hat{h}_{ij}\bigr)^2 
\biggl\{
1- \frac{(M_i-M_j)^2}{m_{\varphi}^2}
\biggr\}
\biggr]
~,
\end{align}
where $\hat{h}_{ij} \equiv \sum_{\alpha,\beta} h_{\alpha \beta}\Omega_{\alpha i}\Omega_{\beta j}$. As it turns out later, the couplings $\hat{h}_{ij}$ are perturbative and $\Gamma_\varphi\ll H_{\rm inf}$ in the parameter region of our interest. We then estimate the reheating temperature as
\begin{align}\label{eq:TR}
T_{R} \simeq \lrf{90}{\pi^{2}g_{\ast}}^{\frac{1}{4}} \sqrt{
\Gamma_{\varphi}M_{P}}~,
\end{align}
where $g_{\ast} = 106.75$ is the relativistic degrees of freedom at the end of reheating. 

If $T_R \lesssim 0.1 M_1$, the produced right-handed neutrinos are out of thermal equilibrium,\footnote{We, however, note that the temperature of the Universe during reheating is in general larger than $T_R$ \cite{Chung:1998rq, Giudice:2000ex} and thus right-handed neutrinos may be produced from the thermal bath even if $T_R \lesssim 0.1 M_1$. In the following analysis, we neglect their contribution just for simplicity. } and their subsequent non-thermal decay generates a lepton asymmetry. In this work, we focus on such a parameter region where this condition is satisfied and leptogenesis takes place non-thermally. For a higher reheating temperature, we need to take account of the inverse decay and scattering processes with the thermal plasma; a detailed analysis for this case will be given on another occasion~\cite{AHNT}.

The baryon asymmetry generated in the non-thermal leptogenesis is computed as~\cite{Lazarides:1991wu, Kumekawa:1994gx, Lazarides:1999dm, Giudice:1999fb, Asaka:1999yd, Asaka:1999jb, Kawasaki:2000ws, Hamaguchi:2002vc}
\begin{align}
Y_{B} \equiv \frac{n_{B}}{s} = 
-\frac{28}{79}  \cdot 
\frac{3T_{R}}{4m_{\varphi}} \cdot (2 \epsilon_{\rm{eff}}) ~,
\label{eq:yb}
\end{align}
where $s$ is the entropy density; the first factor in the right-hand side is the lepton-to-baryon conversion factor via the electroweak sphaleron processes \cite{Harvey:1990qw}; the second one corresponds to the inflaton number per entropy; $\epsilon_{\rm{eff}}$ is the effective asymmetry parameter defined by the averaged asymmetry parameter over the right-handed neutrino decays: 
\begin{align}
\epsilon_{\rm{eff}} &\equiv \frac{1}{2} \sum_{i \leq j} (\epsilon_{i}+\epsilon_{j}) \text{Br}(\varphi \rightarrow N_{i}N_{j})~.
\end{align}
As seen in Eq.~\eqref{eq:yb}, to obtain $Y_B > 0$, we need $\epsilon_{\rm eff} < 0$. 

Now we show the predictions of our model. Let us begin with briefly summarizing the input parameters in this model. As discussed in Sec.~\ref{sec:model}, there are nine parameters in the light neutrino sector, among which four parameters are determined as functions of the other five parameters through the two-zero minor conditions $[{\cal M}_\nu^{-1}]_{\mu\mu} = [{\cal M}_\nu^{-1}]_{\tau\tau} = 0$ as in Eq.~\eqref{eq:m0delalp}. We then fix the remaining five parameters using the neutrino oscillation data as in Eq.~\eqref{eq:oscillation_parameters}. As a result, there is no free parameter in the light neutrino sector. 
For the input parameters in the right-handed neutrino and inflation sectors, we take:
\begin{itemize}
 \item $\lambda$, $\theta$ and $\phi$ in Eq.~\eqref{eq:lamthphi} for the Dirac Yukawa couplings
\item VEV of the U(1)$_{L_\mu-L_\tau}$-breaking Higgs field, $\vev{\sigma}$
\item $\Lambda$ in Eq.~\eqref{eq:lagsigma} 
\end{itemize}
Once the values of $\lambda$, $\theta$ and $\phi$ are chosen, together with the neutrino oscillation parameters~\eqref{eq:oscillation_parameters}, we can uniquely determine the heavy right-handed neutrino mass matrix as discussed in Sec.~\ref{sec:asympar}, as well as the couplings $h_{\alpha \beta}$ for a given value of $\vev{\sigma}$.\footnote{We note that in this case $h_{\alpha \beta } \propto \langle \sigma \rangle^{-1}$. \label{eq:footnoteforh}}
The parameter $\kappa$ in Eq.~\eqref{eq:kappa} and the $e$-folding number $N_e$ are determined by solving Eq.~\eqref{eq:efolding}, Eq.~\eqref{eq:Hinf}, Eq.~\eqref{eq:mphi}, and Eq.~\eqref{eq:TR} for a given set of the above input parameters. Here, we require $N_e\ge 46$ in order to satisfy the constraint on the spectral index $n_s$ within $2\sigma$.
We do not specify the value of the U(1)$_{L_\mu-L_\tau}$ gauge coupling, $g_{Z^\prime}$, as it does not affect the following analysis---we just assume that $g_{Z^\prime}$ is taken to be in the range $\sqrt{\kappa/2} < g_{Z^\prime} \ll (16\pi^2 \kappa)^{1/4}$ to satisfy $m_{\varphi} < 2 m_{Z^\prime}$ and the condition~\eqref{eq:gzllkap}. We can always find such a $g_{Z^\prime}$ for a perturbative value of $\kappa$. 

%%%%%%%%%%%%%%%%%%%%%%%%%%%%%%%%%%%%%%%%%%%%%%%%%%%%%%%
\begin{figure}[t]
\centering
  \includegraphics[width=0.8\columnwidth]{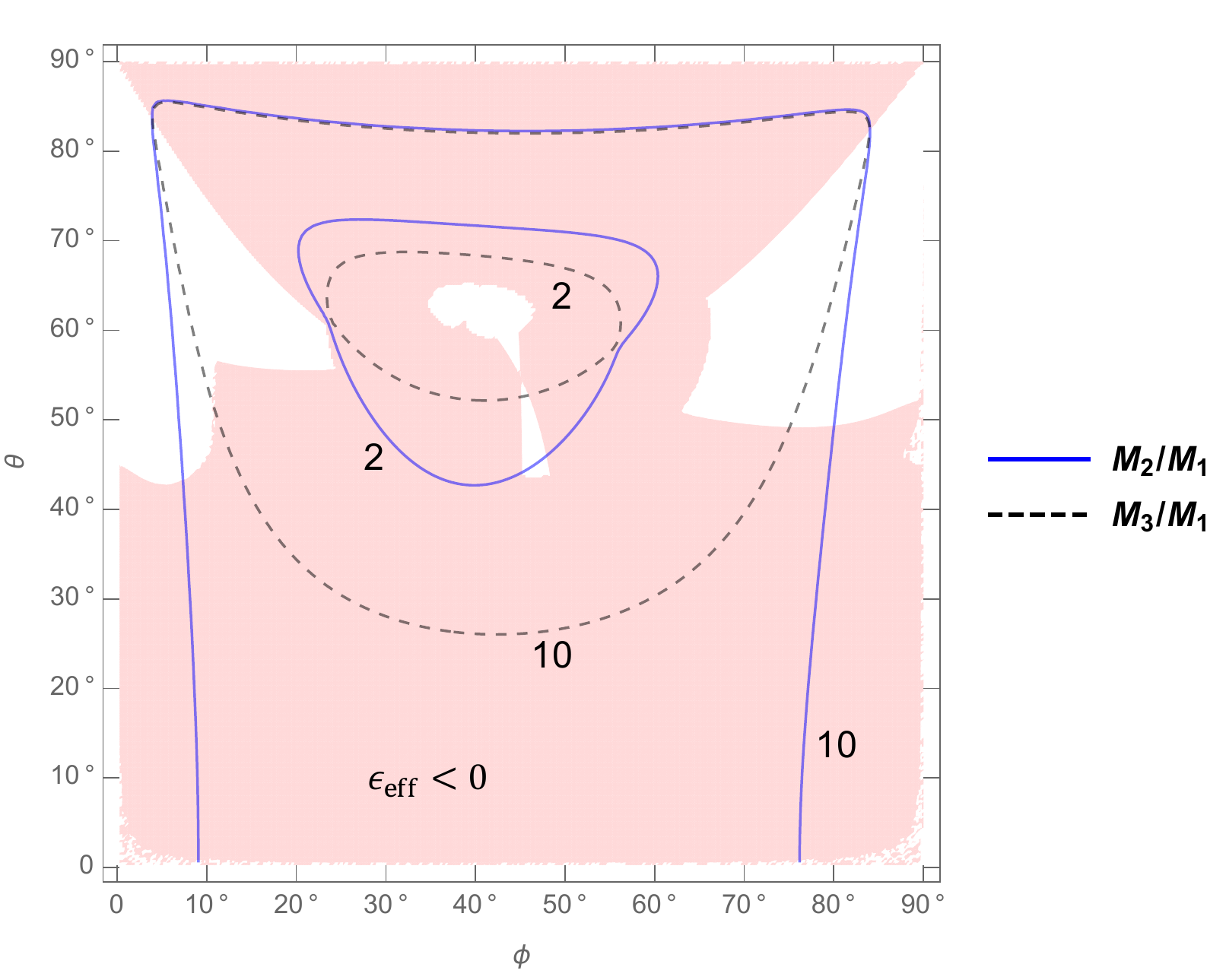}
\caption{The $\phi$-$\theta$ plane for $\lambda = 0.01$, $\vev{\sigma} = 10^{13}$~GeV, and $\Lambda = 10^{16}$~GeV, exhibiting the area where $\epsilon_{\text{eff}} < 0$ in the pink shaded region. The blue solid (gray dashed) contours show the ratio $M_2/M_1$ ($M_3/M_1$). }
\label{fig:eff_epsilon}
\end{figure}
%%%%%%%%%%%%%%%%%%%%%%%%%%%%%%%%%%%%%%%%%%%%%%%%%%%%%%%%

The pink shaded region on the $\phi$-$\theta$ plane in Fig.~\ref{fig:eff_epsilon} shows the area in which $\epsilon_{\text{eff}}$ is predicted to be negative, corresponding to $Y_B > 0$, for $\lambda = 0.01$, $\vev{\sigma} = 10^{13}$~GeV, and $\Lambda = 10^{16}$~GeV. We also show the ratios of the right-handed neutrino masses, $M_2/M_1$ and $M_3/M_1$, by the blue solid and gray dashed contours, respectively, which are identical to the ones shown in Fig.~\ref{fig:epsilon_1}. Comparing Figs.~\ref{fig:epsilon_1} and \ref{fig:eff_epsilon}, we see that the contribution of the heavier right-handed neutrinos to the effective asymmetry parameter is sizable---in a part of the region on the $\phi$-$\theta$ plane, $\epsilon_1 < 0$ but $\epsilon_{\rm eff} > 0$, and vice versa. It is also found that $\epsilon_{\text{eff}} < 0$ is realized in a fairly large fraction of the parameter space.

%%%%%%%%%%%%%%%%%%%%%%%%%%%%%%%%%%%%%%%%%%%%%%%%%%%%%%%%%%%%%%
\begin{figure}[t]
\centering
   \includegraphics[width=0.7\columnwidth]{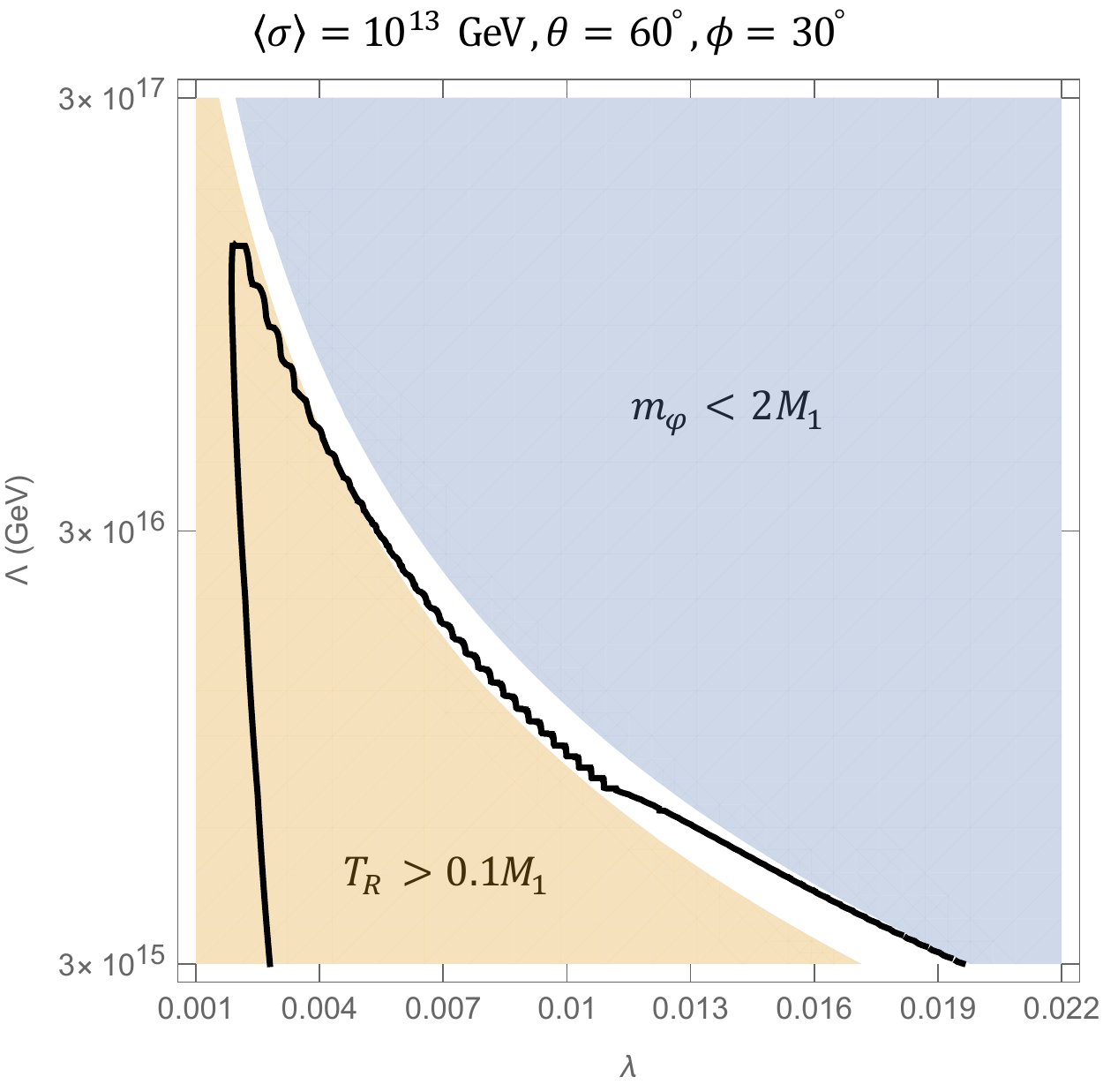}
\caption{The $\lambda$-$\Lambda$ plane for $\theta = 60^{\circ}$, $\phi = 30^{\circ}$, and $\langle \sigma \rangle =10^{13}$~GeV. In the blue shaded region, $m_{\varphi} < 2 M_1$ and thus the decay of inflaton into right-handed neutrinos is kinematically forbidden. In the orange shaded region, $T_R > 0.1 M_1$, for which our analysis for the non-thermal leptogenesis is inappropriate. The black solid curve corresponds to $Y_{B} \simeq 8.7 \times 10^{-11}$~\cite{Aghanim:2018eyx}.  }
\label{fig:baryon_asymmetry}
\end{figure}
%%%%%%%%%%%%%%%%%%%%%%%%%%%%%%%%%%%%%%%%%%%%%%%%%%%%%%%%%%%%%%%

Now we show in Fig.~\ref{fig:baryon_asymmetry} the allowed parameter region of this model on the $\lambda$-$\Lambda$ plane for $\theta = 60^{\circ}$, $\phi = 30^{\circ}$, and $\langle \sigma \rangle =10^{13}$~GeV. This value of $\langle \sigma \rangle$ is chosen such that the cosmic-string bound discussed in Sec.~\ref{sec:inflation} is evaded. In the blue shaded region, $m_{\varphi} < 2 M_1$ and thus the decay of inflaton into right-handed neutrinos is kinematically forbidden. In the orange shaded region, $T_R > 0.1 M_1$, for which our analysis for the non-thermal leptogenesis is inappropriate. 
The black solid curve corresponds to the observed baryon asymmetry $Y_{B} \simeq 8.7 \times 10^{-11}$~\cite{Aghanim:2018eyx}. 
We find that the latter can be reproduced within the allowed parameter region indicated by the white strip between the blue and orange areas. The mass scale of the inflaton and right-handed neutrinos in this case is found to be ${\cal O}(10^{10})~\GeV$, and the reheating temperature is ${\cal O}(10^{8})~\GeV$. Over the parameter space shown in this figure, the couplings $h_{\alpha \beta }$ are ${\cal O} (10^{-3})$ and thus perturbative and compatible with the condition \eqref{eq:hllkap}. The value of the spectral index, $n_s$, is predicted to be $n_s \simeq 0.96$. This prediction can be tested in future CMB experiments such as CMB-S4 \cite{Abazajian:2016yjj, Abazajian:2019eic}.  

The shape of the black solid curve in Fig.~\ref{fig:baryon_asymmetry} can be understood as follows. In the bulk region below this curve, $Y_B$ is predicted to be larger than the observed value. To see the change of $Y_B$ in this region, we first fix $\lambda $ and examine the dependence of $Y_B$ on $\Lambda $. In this case, the right-handed neutrino masses are fixed and thus the couplings $h_{\alpha \beta }$ are also fixed. This means that the inflaton decay width, $\Gamma_\varphi$, is determined solely by the inflaton mass, and approximately goes as $\propto m_{\varphi} \propto \Lambda^{-1}$ in the bulk region. It then follows that $T_R/m_\varphi$, and thus $Y_B$ as well, gets larger for a larger $\Lambda$, roughly scales as $\propto \Lambda^{1/2}$. Just below the blue shaded region, however, the inflaton decay width is highly suppressed by the kinematic factor, resulting in a suppression in the reheating temperature and therefore in $Y_B$. As a result, we can find a correct value of $Y_B$ below the blue shaded area. Next, we fix $\Lambda $ and consider the dependence of $Y_B$ on $\lambda $. In this case, as $\lambda $ decreases, the right-handed neutrino masses, and thus $h_{\alpha \beta }$ as well, get smaller. This leads to a lower reheating temperature. The asymmetric parameters $\epsilon_i$ in Eq.~\eqref{eq:epsilon_1} are also suppressed for a smaller $\lambda$. Hence, $Y_B$ decreases as $\lambda $ gets smaller ($Y_B\sim\lambda^4$) and at a certain point ($\lambda \simeq 0.002$) it coincides with the observed value, $Y_B \simeq 8.7 \times 10^{-11}$~\cite{Aghanim:2018eyx}. 

A large value of $Y_B$ in the bulk region below the black curve would be depleted once we include the thermalization of the right-handed neutrinos and the wash-out of the lepton asymmetry by the thermal bath. This implies that we may find other parameter regions that are compatible with the observed baryon asymmetry, with the thermal effect taken into account. This possibility will be explored in the future~\cite{AHNT}. 

If we take a smaller value of $\langle \sigma \rangle$ than that in Fig.~\ref{fig:baryon_asymmetry}, we need a smaller $\Lambda $ in order to keep $m_\varphi$ larger than $2 M_1$ (see Eq.~\eqref{eq:mphi}). On the other hand, the reheating temperature is larger for a smaller $\langle \sigma \rangle$ since $h_{\alpha \beta }$ increase as $\propto \langle \sigma \rangle^{-1}$, as noted in footnote~\ref{eq:footnoteforh}, and therefore the boundary of the $T_R > 0.1 M_1$ region gets closer to the $m_\varphi < 2 M_1$ region; namely, we need $m_\varphi \simeq 2 M_1$ to suppress the inflaton decay width kinematically so that the non-thermal condition is satisfied. As a result, the allowed parameter region is considerably narrowed down, though the observed value of the baryon asymmetry is still reproduced along the border of the kinematic bound.

All in all, we 
conclude that the non-thermal leptogenesis can be realized successfully in the framework of the minimal gauged U(1)$_{L_\mu-L_\tau}$ model, though the allowed parameter space is rather restricted. Our inflation model can be tested in the future with a precise measurement of $n_s$ in CMB experiments, as well as through the search for the cosmic string signatures in gravitational-wave experiments such as LISA.

%%%%%%%%%%%%%%%%%%%%%%%%%%%%%%%%%%%%
\section{Summary and Discussion}
%%%%%%%%%%%%%%%%%%%%%%%%%%%%%%%%%%%%

We have examined the non-thermal leptogenesis in the framework of the minimal gauged U(1)$_{L_\mu-L_\tau}$ model, where we regard the U(1)$_{L_\mu-L_\tau}$-breaking Higgs field as inflaton. We consider the Lagrangian in Eq.~\eqref{eq:lagsigma} for this field and found that this potential can offer a successful inflation that is consistent with the CMB observation. By requiring that the measured value of the power spectrum be reproduced, we determine the value of the inflaton self coupling $\kappa$ as a function of other input parameters, for which we take $\Lambda$ and $\langle \sigma \rangle$ in Eq.~\eqref{eq:lagsigma}.

As found in the previous studies \cite{Araki:2012ip, Heeck:2014sna, Crivellin:2015lwa, Asai:2017ryy, Asai:2018ocx, Asai:2019ciz}, the light neutrino mass matrix in this model has the two-zero minor structure, which allows us to determine all of the parameters in the light neutrino sector from the neutrino oscillation data. The resultant neutrino mass spectrum and the Dirac CP phase are found to be compatible with the existing experimental bounds~\cite{Asai:2017ryy, Asai:2018ocx, Asai:2019ciz}. The sum of the light neutrino masses and the effective Majorana mass $\langle m_{\beta \beta}\rangle$ predicted in ths model will be tested in future experiments. In addition, the structure of the right-handed neutrino mass matrix is determined by fixing three parameters for the Dirac Yukawa couplings, $\lambda_{\alpha}$ ($\alpha = e, \mu, \tau$)~\cite{Asai:2017ryy}. Our model, therefore, has five free parameters, $\Lambda$, $\langle \sigma \rangle$, and $\lambda_{\alpha}$ ($\alpha = e, \mu, \tau$). 

We then study the non-thermal leptogenesis in our model, focusing on the case where the inflaton decays only into right-handed neutrinos and these right-handed neutrinos are never thermalized after the Universe is reheated. The successive decay of right-handed neutrinos then generates a lepton asymmetry, which is converted to a baryon asymmetry through sphaleron processes. We find that the observed value of baryon asymmetry can be explained in this scenario. In particular, the correct sign of baryon asymmetry can be obtained in a wide range of the parameter space. We recall that our choice of $\delta > \pi $, which is favored by the present neutrino oscillation data~\cite{Esteban:2018azc, NuFIT4_1, Abe:2019vii}, was crucial in obtaining this result; if we instead chose $\delta < \pi $, we would obtain a wrong sign for the baryon asymmetry in most parameter regions.

Our analysis shows that baryon asymmetry tends to be overproduced in the non-thermal leptogenesis scenario. This observation gives a strong motivation for a more detailed study on leptogenesis in this model with the effect of the thermal plasma taken into account---we shall return to this issue in future work~\cite{AHNT}.

%%%%%%%%%%%%%%%%%%%%%%%%%%%%%
\section*{Acknowledgements}
%%%%%%%%%%%%%%%%%%%%%%%%%%%%%

This work is supported in part by 
JSPS KAKENHI (No. JP19J13812 [KA])
and the Grant-in-Aid for 
Innovative Areas (No.19H05810 [KH], No.19H05802 [KH], No.18H05542 [NN]),
Scientific Research B (No.20H01897 [KH and NN]), and Young Scientists B (No.17K14270 [NN]).

%%%%%%%%%%%%%%%%%%%%%%%%%

%%%%%%%%%%%%%%%%%%%%%%%%%%%%%%%%%%%%%%%%%%%%%%%%%%

{
\bibliographystyle{utphysmod}
\bibliography{mu-tau-LG} %
}
\end{document}